\documentclass[journal]{IEEEtran}
\usepackage{graphicx}
\usepackage[cmex10]{amsmath}
\usepackage{cases}
\usepackage{amsthm}
\usepackage{cite}
\usepackage{amssymb}
\usepackage{algorithm}
\usepackage{algorithmic}
\usepackage{multirow}
\usepackage{amsmath}
\usepackage{xcolor}
\usepackage{subeqnarray}
\usepackage{cases}
\usepackage{enumerate}
\usepackage{cuted}
\usepackage{booktabs}
\usepackage{mathrsfs}
\usepackage{bbm}
\usepackage{graphicx}

\ifCLASSINFOpdf
\else
\fi

\begin{document}
\title{Frequency Diverse RIS (FD-RIS) Enhanced Wireless Communications via Joint Distance-Angle Beamforming}
\author{Han Xiao,~\IEEEmembership{Student Member,~IEEE,} Xiaoyan Hu$^*$,~\IEEEmembership{Member,~IEEE,} Wenjie Wang,~\IEEEmembership{Senior Member,~IEEE,} \\
Kai-Kit~Wong,~\IEEEmembership{Fellow,~IEEE}, Kun~Yang,~\IEEEmembership{Fellow,~IEEE}
	
\thanks{H. Xiao, X. Hu, and W. Wang are with the School of Information and Communications Engineering, Xi'an Jiaotong University, Xi'an 710049, China. (email: hanxiaonuli@stu.xjtu.edu.cn, xiaoyanhu@xjtu.edu.cn, wjwang@mail.xjtu.edu.cn. \emph{(Corresponding author: Xiaoyan Hu.)}}
\thanks{K.-K. Wong is with the Department of Electronic and Electrical Engineering, University College London, London WC1E 7JE, U.K. (email: kai-kit.wong@ucl.ac.uk).}
\thanks{K. Yang is with the School of Computer Science and Electronic Engineering, University of Essex, Colchester CO4 3SQ, U.K. (e-mail: kunyang@ essex.ac.uk).}
}

\maketitle

\begin{abstract}
The conventional reconfigurable intelligent surface (RIS) assisted far-field communication systems can only implement angle beamforming, which actually limits the capability for reconfiguring the wireless propagation environment. To overcome this limitation, this paper proposes a newly designed frequency diverse RIS (FD-RIS), which can achieve joint distance-angle beamforming with the assistance of the time modulation technology. The signal processing model for FD-RIS-aided wireless communications is first derived. Then, an optimization problem aimed at maximizing the achievable rate is formulated where the frequency-time modulations are jointly optimized to achieve distance-angle beamforming. Furthermore, a novel iterative algorithm based on the cross-entropy optimization (CEO) framework is proposed to effectively handle the non-convex optimization problem. The numerical results validate that the proposed FD-RIS assisted communication scheme can achieve a notable performance improvement compared with the baseline scheme utilizing traditional RIS. In addition, the effectiveness of the proposed CEO algorithm is further verified by comparing with the benchmark using the genetic algorithm (GA).
\end{abstract}
\begin{IEEEkeywords}
Frequency diverse RIS (FD-RIS), time modulation, distance-angle beamforming, cross-entropy optimization (CEO)
\end{IEEEkeywords}
\IEEEpeerreviewmaketitle

\section{Introduction}\label{sec:S1}
Currently, the spectral efficiency of mobile communication systems is steadily approaching its limitation. ``How to further improve the spectral efficiency and system capacity?'' is the key challenge that the next-generation wireless communication technologies need to address. As a pivotal technology of the 5G and  6G, massive multiple-input multiple-output (MIMO) technique exhibits the enormous potential in enhancing spectral efficiency, security, and reliable connectivity \cite{andrews2014will}. However, as the size of antenna arrays continues to expand, the MIMO systems based on phased arrays will encounter numerous challenges such as high costs, high cimplexity, high power consumption and significant insertion losses, which will hinder the large-scale deployment of massive MIMO systems in 5G and 6G networks \cite{shlezinger2021dynamic}.
The technology of reconfigurable intelligent surface (RIS) offers a cost-effective and hardware-efficient solution to overcome the obstacles encountered by the massive MIMO systems. Actually, RIS as an affordable two-dimensional metasurface  has the capability to control the electromagnetic properties of the incident signals \cite{hu2022irs}.
It is verified that RIS exhibits significant application potentials and has been incorporated into various communication systems  for performance enhancement \cite{Han2024simultaneously}.

In fact, even with the assistance of RIS, the enhancement of system capacity for far-field communications is still severely limited, primarily due to the following reasons: (\romannumeral 1)  For far-field communications, the signal undergoes large-scale path loss during transmissions especially with blockages.
Although RIS can reconfigure the propagation environment, energy loss remains an unavoidable factor. 
(\romannumeral 2) RIS can only manipulate the angle-dependent beamforming in far-field communication scenarios, and lacks the capability to directly control the distance-dependent beamforming compared to near-field communication scenarios \cite{shen2023multi}.
Based on the above reasons, it is important to improve the energy efficiency of far-field communications, so as to further augment the system capacity.
Actually, by introducing the control of distance dimension for incident signals is highly beneficial for  RIS's passive beamforming.
However, ``How to integrate the distance dimension into RIS's beamforming design? " poses a significant challenge.

Fortunately, the frequency diverse array (FDA) exemplifies effective distance-angle beamforming control for the far-field communications \cite{antonik2006frequency}. This is achieved by applying frequency offsets across the array elements, enabling the array to produce a coupled distance-angle pattern.
Inspired by the intrinsic mechanism of the FDA antennas, it is possible for RIS to generate joint distance-angle beamforming if it can incorporate frequency diversity into the incident signals. Nevertheless, unlike FDA antennas, RIS as a passive device does not have the capability to achieve signal frequency diversity through radio frequency (RF) links.
Authors in \cite{zhang2018space} indicate that by implementing time modulation on RIS, the incident signal can be dispersed into a series of harmonic signals, each of which has a frequency offset from the central carrier frequency.
Therefore, RIS with time modulation may be capable of enabling frequency diversity for the incident signals and further achieve joint distance-angle beamforming.

In this paper, we name the newly designed RIS with the function of frequency diversity as ``Frequency Diverse RIS (FD-RIS)". Although it is possible for FD-RIS to achieve the distance-angle control for the incoming signals, the systematic analytical framework for FD-RIS is unclear which may hinder its applications in wireless communication systems. Hence, the contributions of this paper are summarized as below:
\begin{itemize}
  \item We initially  establish the general signal processing model and the equivalent channel model of the FD-RIS assisted communication systems. 
  \item The system capacity maximization problem is formulated, and we design an effective algorithm based on the cross-entropy optimization (CEO) framework.
  \item The effectiveness of the designed FD-RIS as well as the proposed algorithm is verified through simulation results.
\end{itemize}

\section{Signal Processing Model for FD-RIS}\label{sec:S2}
In this section, we first illustrate the rational of the FD-RIS and derive its signal processing model.
The time modulation technology introduced in \cite{zhang2018space}  is essential for implementing the function of FD-RIS, through which the reflected signals from the FD-RIS have the property of  frequency diversity as the FDA, and thus can achieve distance-angle beamforming.

First, we introduce a periodic square wave function $V(t)$ with period $T_0$ and its expression in the $g$-th ($g\in \mathbb{Z}$) period, i.e.,  $t\in[gT_0,(g+1)T_0]$,  is given by
\begin{align}\label{Vmn}
	V_g(t)=
	\begin{cases}
1,&gT_0<t\leq gT_0+\tau,\\
0,&gT_0+\tau<t\leq (g+1)\tau.
	\end{cases}
\end{align}
It is assumed that the FD-RIS is installed with  $S=M\times N$ passive elements with indexes $m\in\mathcal{M}\triangleq\{1,\cdots,M\}$, $n\in\mathcal{N}\triangleq\{1,\cdots,N\}$. For the $(m, n)$-th element of the FD-RIS, the time-modulated reflection coefficient can be expressed as
\begin{align}\label{Rmn}
    \Upsilon_{mn}(t)=& \sum\limits_{l=1}^{L}\Upsilon^l_{mn} V(t-(l-1)\tau),
\end{align}
where $\Upsilon^l_{mn}$ denotes the reflection coefficient in the $l$-th time slot of each time modulation period $T_0$, 
with slot length $\tau=\frac{T_0}{L}$ and $l\in\mathcal{A}\triangleq\{1,\cdots,L\}$. Here, $L$ is  the length of time modulation sequence coding within a period.

According to the Fourier series expansion theory of the periodic function, $V(t-(l-1)\tau)$ can be derived as
\begin{align}\label{Vmnl}
		V(t-(l-1)\tau)=\sum\limits_{z=-\infty}^{z=+\infty}a_{lz}e^{j2\pi zf_0t},
\end{align}
where $a_{lz}=e^{-j2\pi z\frac{l-1}{L}}\frac{1}{L}\mathrm{sinc}(\frac{\pi z}{L})e^{\frac{-j\pi z}{L}}$ with $\mathrm{sinc}(x)=\frac{\sin x}{x}$ is the Fourier series coefficients and $f_0=\frac{1}{T_0}$ is the time-modulated frequency.

Next, we will describe the signal processing model for the FD-RIS.
Assuming that the signal emitted by the base station (BS) is denoted as $\sqrt{P}s(t)e^{j2\pi f_\mathrm{c}t}$, where $P$ is the transmitting power, $s(t)$ indicates the narrow-band complex envelope, and $f_c$ represents the carrier frequency. Once the transmitted signal approach the FD-RIS, the reflected signal  can be expressed as
\begin{itemize}
     \item $x(t)=\sqrt{P}s(t)\eta(d_\mathrm{br})\sum\limits_{m=1}^{M}\sum\limits_{n=1}^{N} e^{j2\pi f_\mathrm{c}(t-\frac{d_\mathrm{br}^{mn}}{c})}\Upsilon_{mn}(t),$
 \end{itemize}
where $\eta(d_\mathrm{br})$ represents the large-scale path loss of the signal with $d_\mathrm{br}$ being the distance between the BS and the FD-RIS.
Also, $d_\mathrm{br}^{mn}$ is the spatial distance between the BS and the $(m, n)$-th element of FD-RIS, given as $d_\mathrm{br}^{mn}=d_\mathrm{br}-(m-1)d\sin(\theta_\mathrm{br})\cos(\phi_\mathrm{br})-(n-1)d\sin(\theta_\mathrm{br})\sin(\phi_\mathrm{br})$,
where $\theta_\mathrm{br}$ and $\phi_\mathrm{br}$ are the elevation and azimuth angles of arrival from the BS to the FD-RIS,  $d$ is the distance between two adjacent elements of the FD-RIS, and $c$ is the speed of light.

According to the derivations in \eqref{Rmn}-\eqref{Vmnl}, the signal reflected by the FD-RIS can be further modeled as
\begin{itemize}\label{eq_signal_model}
	\item $ x(t) =\sqrt{P}s(t)\eta(d_\mathrm{br})\sum_{m=1}^{M}\sum_{n=1}^{N} e^{-j2\pi f_\mathrm{c}\frac{d_\mathrm{br}^{mn}}{c}}\sum_{z=-\infty}^{z=+\infty}\\\sum_{l=1}^{L}\Upsilon^l_{mn}a_{lz}e^{j2\pi (f_\mathrm{c}+zf_0)t},$
\end{itemize}
from which we can observe that the FD-RIS can transform the incident signal into a series of harmonic signals with frequency $(f_\mathrm{c}+zf_0), z\in(-\infty,+\infty) $. Hence, the FD-RIS introduces frequency diversity for the incident signal, indicating that it has the ability to achieve the distance-angle coupling beamforming by exploring the concept of the FDA antenna.

Based on the expression of Fourier series coefficients $a_{lz}$, it is easy to note that the maximum amplitude of the the harmonic signals  decreases as the harmonic order $|z|$ increases. To facilitate the practical manipulations, we can utilize the set of harmonic signals with orders in $z\in[-Z,Z]$ to approximately represent the reflected signal.
\section{System model and Problem formulation}\label{sec:S3}
\subsection{System Model}
\begin{figure}[ht]
	\centering
	\includegraphics[scale=0.35]{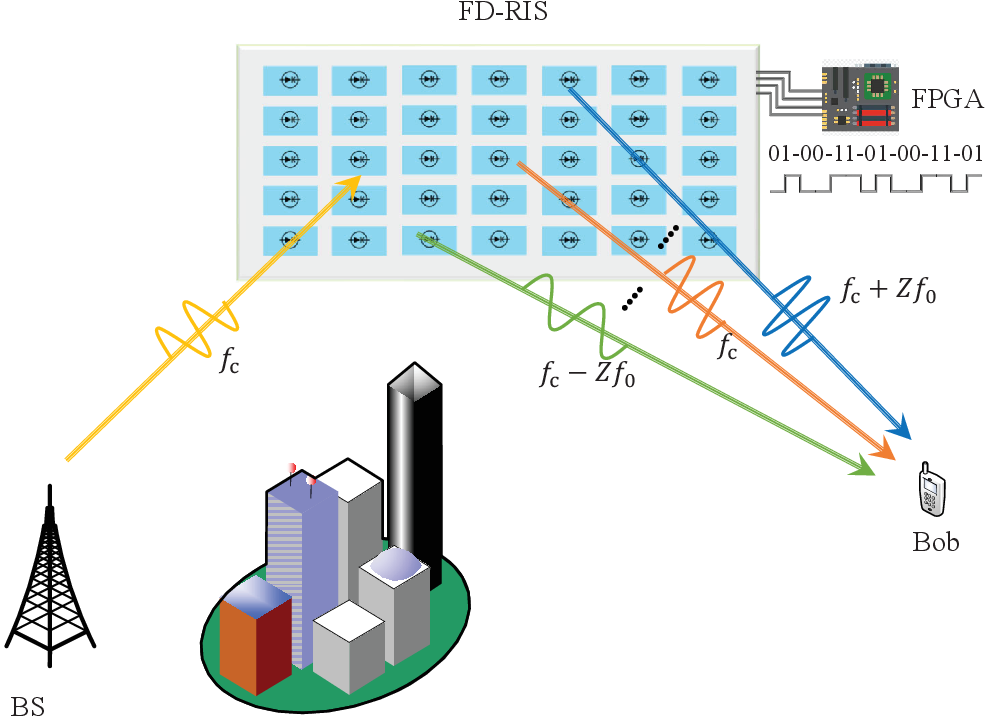}\\
	\caption{The system model of FD-RIS-assisted wireless networks.}\label{fig:scenario}
\end{figure}

Fig. \ref{fig:scenario} shows a basic system model of the FD-RIS-assisted wireless communication networks, which comprises a single-antenna BS, a single-antenna legitimate user (Bob), and a FD-RIS with $S=M\times N$ elements. It is assumed that the line of sight (LoS) connection between the BS and the user is obstructed by buildings or trees, which commonly occurs  in far-filed urban communication scenarios.

On the basis of the analysis above, the received signal at the user located at $(d_\mathrm{ru},\theta_\mathrm{u},\phi_\mathrm{u})$ can be expressed as
\begin{align}\label{eq_y1}
	y=\eta(d_\mathrm{ru})x\big(t-\frac{d_\mathrm{ru}^{mn}}{c}\big)+n, 
\end{align}
where
$\eta(d_\mathrm{ru})$ is the large-scale path loss of the signal from the FD-RIS to the user with spatial distance $d_\mathrm{ru}$. $n\sim \mathcal{CN}(0, \sigma^2)$ represents the additive noise with  $\sigma^2$ being the noise power. In addition, $d_\mathrm{ru}^{mn}$ indicates the distance between the user and the $(m, n)$-th element of the FD-RIS, which is given by
$d_\mathrm{ru}^{mn}=d_\mathrm{ru}-(m-1)d\sin(\theta_\mathrm{ru})\cos(\phi_\mathrm{ru})
-(n-1)d\sin(\theta_\mathrm{ru})\sin(\phi_\mathrm{ru})$,
with $\theta_\mathrm{ru}$ and $\phi_\mathrm{ru}$ being the elevation and azimuth angles of departure from the FD-RIS to the user.
\textcolor{blue}{Thus, the energy of the received signal is given by
	\begin{itemize}
		\item $\left|y\right|^2=P\eta^2(d_\mathrm{br})\eta^2(d_\mathrm{ru})\big|\sum_{m=1}^{M}\sum_{n=1}^{N} e^{j2\pi f_\mathrm{c}\frac{\Gamma_\mathrm{br}^{mn}}{c}}\\\sum_{z=-Z}^{z=+Z}\sum_{l=1}^{L}\Upsilon^l_{mn}a_{lz}e^{j2\pi zf_0(t-\frac{d_\mathrm{ru}-\Gamma_\mathrm{ru}^{mn}}{c})}\big|^2+\sigma^2,$
	\end{itemize}
where $\Gamma_\mathrm{br}^{mn}=(m-1)d\sin(\theta_\mathrm{br})\cos(\phi_\mathrm{br})+(n-1)d\sin(\theta_\mathrm{br})$ $\sin(\phi_\mathrm{br})$, $\Gamma_\mathrm{ru}^{mn}=(m-1)d\sin(\theta_\mathrm{ru})\cos(\phi_\mathrm{ru})+(n-1)d$ $\sin(\theta_\mathrm{ru})\sin(\phi_\mathrm{ru})$.
	From the expression of the received signal energy, we observe that due to the presence of frequency diversity, the energy of the received signal is influenced not only by the angles and  but also by the distance between the FD-RIS and the user. This means that we are able to appropriately design the reflection coefficients of the FD-RIS in order to effectively control the signal energy focusing towards the intended location from angle and distance dimensions, thereby enhancing the SINR of the desired position. }

To facilitate the further operations, \eqref{eq_y1} can be equivalently re-expressed in the form with matrix manipulation as
\begin{align}
	y=\mathbf{h}_\mathrm{br}^H\boldsymbol{\Theta}\mathbf{h}_\mathrm{ru}s(t)+n,
\end{align}
where
\begin{itemize}
	\item $\mathbf{h}_\mathrm{br}=\sqrt{P}\eta(d_\mathrm{br})e^{j2\pi f_\mathrm{c}t}[e^{-j2\pi f_\mathrm{c}\frac{d_\mathrm{br}}{c}}, \cdots, e^{-j2\pi f_\mathrm{c}\frac{d_\mathrm{br}-\Gamma_\mathrm{br}^{mn}}{c}},\\ \cdots, e^{-j2\pi f_\mathrm{c}\frac{d_\mathrm{br}-\Gamma_\mathrm{br}^{MN}}{c}}]^T
$		\vspace{1mm}
\item  $\mathbf{h}_\mathrm{ru}=\eta(d_\mathrm{ru})[e^{-j2\pi f_\mathrm{c}\frac{d_\mathrm{ru}}{c}}, \cdots, e^{-j2\pi f_\mathrm{c}\frac{d_\mathrm{br}-\Gamma_\mathrm{ru}^{mn}}{c}}, \cdots, \\e^{-j2\pi f_\mathrm{c}\frac{d_\mathrm{br}-\Gamma_\mathrm{ru}^{MN}}{c}}]^T
$
\vspace{1mm}
	\item $\boldsymbol{\Theta}=\operatorname{Diag}\{\theta_1,\cdots, \theta_s,\cdots, \theta_{S}\}$ with $s=1,\cdots,S$,
	\item $ \theta_s=\sum_{z=-Z}^{z=+Z}\sum_{l=1}^{L}\Upsilon^l_{mn}a_{lz}e^{j2\pi zf_0(t-\frac{d^{mn}_\mathrm{ru}}{c})}=\boldsymbol{\gamma}_{s}^T\mathbf{A}_{s}\mathbf{b}_{s},$
\end{itemize}
$\boldsymbol{\gamma}_{s}=[\Upsilon^1_{mn}, \cdots, \Upsilon^l_{mn}, \cdots, \Upsilon^L_{mn}]^T$, the element of $\mathbf{A}_{s}\in \mathbb{C}^{L\times(2Z+1)}$ is given as $\mathbf{A}_{s}(l, z+Z+1)=a_{lz}$, and
 $\mathbf{b}_{s}=[e^{-j2\pi Zf_0(t-\frac{d^{mn}_\mathrm{ru}}{c})}, \cdots, 1, \cdots, e^{j2\pi Zf_0(t-\frac{d^{mn}_\mathrm{ru}}{c})}]^T$.

In this way, we construct an equivalent channel $\mathbf{h}_\mathrm{br}^H\boldsymbol{\Theta}\mathbf{h}_\mathrm{ru}$ similar to the traditional RIS,  where $\mathbf{h}_\mathrm{br}$ and $\mathbf{h}_\mathrm{ru}$ respectively represent the equivalent channels between the BS and the FD-RIS as well as between the  FD-RIS and the user, and $\boldsymbol{\Theta}$ is the equivalent reflection coefficient matrix of the FD-RIS.
Hence, the achievable communication capacity of the system can be expressed as
 $R=\log_2\left(1+\left|\mathbf{h}_\mathrm{br}^H\boldsymbol{\Theta}\mathbf{h}_\mathrm{ru}\right|^2/\sigma^2\right).$
\subsection{Problem Formulation }
In order to assess the capability of the FD-RIS, 
we formulate an optimization problem to maximize the achievable communication rate as below
\begin{subequations}\label{eq_orig_opti}
	\begin{align}
	&\max _{\boldsymbol{\Upsilon}, f_0}~~ R,\notag \\
	&\text { s.t. }~f_{\min}\leq f_0\leq f_{\max},\label{eq_orig_opti_1}\\
	&\qquad~\Upsilon^l_{mn}\in\mathcal{G},~m\in\mathcal{M},~n\in\mathcal{N},~l\in\mathcal{A},\label{eq_orig_opti_2}
	\end{align}
\end{subequations}
where $f_{\min}$ and $f_{\max}$ indicate the lower and upper bounds of the time modulation frequency $f_0$. 
It is assumed that the FD-RIS possesses the $b$-bit phase shift resolution for each reflection coefficient $\Upsilon^l_{mn}$, and thus $\mathcal{G}\triangleq\{\varphi_q=e^{jq\frac{2\pi}{2^{b}}}| q=1,\cdots, Q\},~ Q=2^{b}$.
Through jointly optimizing $f_0$ and $\boldsymbol{\Upsilon}=\{\Upsilon^l_{mn}\}$,
the FD-RIS's equivalent reflection coefficient matrix, i.e., $\boldsymbol{\Theta}$, can be effectively designed.
Actually, the optimization problem \eqref{eq_orig_opti} is difficult to solve, due to the non-convex objective function and the discrete variable $\Upsilon^l_{mn}$.
Next, we propose an optimization algorithm based on the CEO framework to address these challenges.
\section{Algorithm Design}\label{sec:S4}
The CEO framework is a probability-based learning technique utilized to solve complex  problems \cite{rubinstein2004cross}. Its fundamental idea can be summarized as: (\romannumeral 1)   Producing potential solutions from the sampling distribution characterized by the tilting parameters, and evaluating these solutions by calculating the corresponding objective function value.
(\romannumeral 2)  Updating the tilting parameters of the sampling distribution involves minimizing the cross-entropy between the current distribution and reference distribution derived from these highly effective solutions;  (\romannumeral 3)  Continuously generating new samples based on updated parameters until the difference in objective function values between iterations meets a set threshold.

\vspace{-2mm}
\subsection{Design for sampling distribution}
To effectively address the optimization problem using the CEO framework, it is imperative to initially establish two suitable sampling distribution models to generate the discrete variable $\Upsilon_{mn}^l$ and the continuous variable $f_0$, respectively.
\subsubsection{Sampling distribution model for the reflection coefficients in $\boldsymbol{\Upsilon}$}
Let $\boldsymbol{\gamma}=\{\gamma_p\}_{p=1}^{P=SL}=[\boldsymbol{\gamma}_1^T, \cdots, \boldsymbol{\gamma}_s^T, \cdots, \boldsymbol{\gamma}_{S}^T ]^T\in\mathbb{C}^{P\times 1}$. Note that each entry in $\boldsymbol{\gamma}$ can be mapped into a specific reflection coefficient in $\boldsymbol{\Upsilon}$ effortlessly.
To construct a stochastic vector $\boldsymbol{\gamma}$, we can sample each $\gamma_p$ independently from the obtained or pre-determined discrete probability distribution $[P_{p1}, \cdots, P_{pq}, \cdots, P_{pQ}]$ comprising of $Q$ probability values where $P_{pq}=\operatorname{Pr}(\mathcal{I}_p(\boldsymbol{\gamma})=\varphi_q)$ represents the likelihood of the $p$-th entry in  $\boldsymbol{\gamma}$ selecting the $q$-th element in $\mathcal{G}$, and satisfies $\sum_{q=1}^{Q}P_{pq}=1$. A matrix denoted as $\mathbf{P}=[P_{pq}]\in\mathbb{R}^{P\times Q}$ can be constructed by aggregating all probabilities. Therefore, the joint sampling distribution of $\boldsymbol{\gamma}$ can be parametrized by $\mathbf{P}$, which is given as
\begin{align}
\mathscr{F}(\boldsymbol{\gamma}; \mathbf{P})=\prod_{p=1}^{P}\sum_{q=1}^{Q}P_{pq}\mathbbm{1}_{\{\mathcal{I}_p(\boldsymbol{\gamma})=\varphi_q\}},
\end{align}
where $\mathbbm{1}_{\{\cdot\}}$ represents the indicator function for a event, and $\mathcal{I}_p(\mathbf{a})$ denotes the $p$-th entry of the vector $\mathbf{a}$.
\subsubsection{Sampling distribution model for the modulation frequency $f_0$}
Different from the reflection coefficients, the modulation frequency $f_0$ is a continuous variable. Here, the Gaussian distribution is a prevalent and optimal selection for generating random modulation frequency $f_0$ \cite{chen2024designing}. Hence, the sampling distribution $f_0$ can be expressed as
$	\mathscr{F}(f_0; \tilde{\sigma};\mu )=\frac{1}{\sqrt{2\pi}\tilde{\sigma}}e^{-\frac{(f_0-\mu)^2}{2\tilde{\sigma}^2}},$
where $\tilde{\sigma}^2$ and $\mu$ represent the variance and mean of the Gaussian sampling distribution, respectively. 

\subsubsection{Joint sampling distribution model}  According to the analysis above, the joint sampling distribution for handling optimization problem \eqref{eq_orig_opti} can be expressed as
\begin{align}\label{eq_joint_sd}
	\mathscr{F}(\boldsymbol{\Xi};\boldsymbol{\rho})=\mathscr{F}(\boldsymbol{\gamma}; \mathbf{P})\times  \mathscr{F}(f_0; \tilde{\sigma};\mu ),
\end{align}
where $\boldsymbol{\rho}\triangleq\{\mathbf{P}, \tilde{\sigma},\mu \}$ denotes the tilting parameter set of the joint sampling distribution, and the sampling solution denoted by $\boldsymbol{\Xi}\triangleq\{\boldsymbol{\gamma}, f_0\}$ can be  generated from $\mathscr{F}(\boldsymbol{\Xi};\boldsymbol{\rho})$.

\subsection{Updating Formulas for Tilting Parameters}
As mentioned before, the CEO framework updates the tilting parameters by minimizing the cross-entropy between the current sampling distribution and the reference sampling distribution derived from the top-performing feasible solutions. Specifically, we first produce $K$ feasible samples $\{\boldsymbol{\Xi}_k\}_{k=1}^K$ based on the joint sampling distribution in \eqref{eq_joint_sd}, and then calculate the objective function value $R(\boldsymbol{\Xi}_k)$ of each candidate. To identify these highly effective solutions, the generated candidates $\{\boldsymbol{\Xi}_k\}_{k=1}^K$ are ranked by their corresponding objective function value in the descending order. The candidate that possesses the $k$-th largest objective function value is re-denoted as $\boldsymbol{\Xi}_{[k]}$, which means $R(\boldsymbol{\Xi}_{[1]})\geq R(\boldsymbol{\Xi}_{[2]})\geq \cdots \geq R(\boldsymbol{\Xi}_{[k]}), \cdots, \geq R(\boldsymbol{\Xi}_{[K]})$. The first $K^\mathrm{elite}=\varrho K$ samples are selected to form the elite set, denoted by $\mathcal{S}=\{\boldsymbol{\Xi}_{[k]}\}_{k=1}^{K^\mathrm{elite}}$, where $\varrho\in(0,1)$ indicates the percentage of the top-performing samples chosen for the elite set.

The selected top-tier candidates will be employed to establish the reference distribution, and the tilting parameters will be adjusted by the obtained solutions that minimizing the cross-entropy between the present distribution $\mathscr{F}(\cdot;\boldsymbol{\rho})$ and the reference distribution. As presented in \cite{rubinstein2004cross}, the cross-entropy optimization problem can be equivalently transformed as
\begin{subequations}\label{eq_entropy}
	\begin{align}
		&\max _{\boldsymbol{\rho}}~~ \frac{1}{K}\sum_{k=1}^{K^\mathrm{elite}}\ln\mathscr{F}(\boldsymbol{\Xi}_{[k]};\boldsymbol{\rho}),\notag \\
		&~\text { s.t. }~\sum_{q=1}^{Q}P_{pq}=1,~\forall p\in\mathcal{P}\triangleq\{1, \cdots, P\}.\label{eq_entropy_1}
	\end{align}
\end{subequations}
which is non-convex with mixed discrete and continuous optimization parameters.
Actually, the Lagrange multiplier method can be leveraged to effectively handle the optimization problem \eqref{eq_entropy}. To utilize this approach, we first construct the Lagrange function of problem \eqref{eq_entropy}, which is given by
\begin{itemize}
\item $	\mathcal{L}=\frac{1}{K}\sum_{k=1}^{K^\mathrm{elite}}\ln\mathscr{F}(\boldsymbol{\Xi}_{[k]};\boldsymbol{\rho})+ \sum_{p=1}^{P}\epsilon_p(\sum_{q=1}^{Q}P_{pq}-1),$
\end{itemize}
where $\epsilon_p$ is the Lagrange multiplier associated with the equality constraint in \eqref{eq_entropy_1}. The first-order partial derivative  of the Lagrange function versus $\mathbf{P}$, $\tilde{\sigma}$ and $\mu$ can be derived as
\textcolor{blue}{
\begin{itemize}
	\item $\frac{\partial \mathcal{L}}{\partial P_{pq}}=\frac{1}{K}\sum_{k=1}^{K^\mathrm{elite}}\frac{\mathbbm{1}_{\{\mathcal{I}_p(\boldsymbol{\gamma}_{[k]})=\varphi_q\}}}{P_{pq}}+\epsilon_p,$ $\forall p\in\mathcal{P}$,
	\vspace{1mm}
	\item $\frac{\partial \mathcal{L}}{\partial\tilde{\sigma}}=\frac{1}{K}\sum_{k=1}^{K^\mathrm{elite}}\frac{((f_0)_{[k]}-\mu)^2-K^\mathrm{elite}\tilde{\sigma}^2}{\tilde{\sigma}^3},$
		\vspace{1mm}
	\item $\frac{\partial \mathcal{L}}{\partial\mu}=\frac{1}{K}\sum_{k=1}^{K^\mathrm{elite}}\frac{(f_0)_{[k]}-K^\mathrm{elite}\mu}{\tilde{\sigma}^2}$.
\end{itemize}
Let $\frac{\partial \mathcal{L}}{\partial P_{pq}}=0$, $\frac{\partial \mathcal{L}}{\partial \tilde{\sigma}}=0$ and $\frac{\partial \mathcal{L}}{\partial\mu}=0$,} we have the closed-form solutions given below
\begin{itemize}
	\item $P_{pq}=-\frac{\sum_{k=1}^{K^\mathrm{elite}}\mathbbm{1}_{\{\mathcal{I}_p(\boldsymbol{\gamma}_{[k]})=\varphi_q\}}}{K\epsilon_p}, ~\epsilon_p=-\frac{K^\mathrm{elite}}{K},$
		\vspace{1mm}
	\item $\tilde{\sigma}=\sqrt{\frac{\sum_{k=1}^{K^\mathrm{elite}}((f_0)_{[k]}-\mu)^2}{K^\mathrm{elite}}},~ \mu=\frac{\sum_{k=1}^{K^\mathrm{elite}}(f_0)_{[k]}}{K^\mathrm{elite}}$.	
\end{itemize}

To mitigate the risk of converging towards local optimal solutions for the CEO-based algorithm, here a smoothing method commonly used in reinforcement learning is applied during successive iterations to adjust the tilting parameters. \textcolor{blue}{Hence, the tilting parameters for the $i$-th iteration are modified according to the subsequent approach:
\begin{align}
	&P_{pq}^{(i)}\leftarrow\xi P_{pq}^{(i)}+(1-\xi)P_{pq}^{(i-1)},\\
	&\tilde{\sigma}^{(i)}\leftarrow\xi \tilde{\sigma}^{(i)}+(1-\xi)\tilde{\sigma}^{(i-1)},\\
	&\mu^{(i)}\leftarrow\xi \mu^{(i)}+(1-\xi)\mu^{(i-1)},
\end{align}
where $\xi\in(0,1)$ denotes the smoothing parameter, $\leftarrow$ denotes the assignment operation.}  $K$ novel feasible candidates will be generated utilizing the revised tilting parameters, followed by the selection of elite solutions for updating the tilting parameters in the next operations.
\textcolor{blue}{The proposed algorithm faces a challenge in evaluating the objective values of $K$ feasible solutions, resulting in a computing complexity of $\mathcal{O}(LK)$, where $L$ is the number of iterations. Notably, the algorithm's computational complexity is much lower compared to various convex optimization algorithms.}
\section{Numerical Results}\label{sec:S5}
To evaluate the potentials of the designed FD-RIS in improving the system capacity and the effectiveness of the proposed CEO algorithm (FD-RIS CEO), the numerical simulations are implemented in comparison with two general benchmark schemes, i.e., the genetic algorithm (GA) scheme (FD-RIS GA), the traditional RIS-aided scheme with the proposed CEO method  (RIS CEO) and the GA method (RIS GA). We consider the following parameters: $Z=3$, $L=7$, $f_\mathrm{c}=28$ GHz, $f_{\min}=100$ KHz, $f_{\max}=280$ KHz, $b=2$, $\sigma^2=-110$ dBm, $\xi=0.65$ and $d_\mathrm{br}=30$ m. The large-scale path-loss model $\eta(d)=-30-22\log d$ dB \cite{Xiaostar-ris2024} is leveraged.
\begin{figure}[h]
\centering
\includegraphics[scale=0.37]{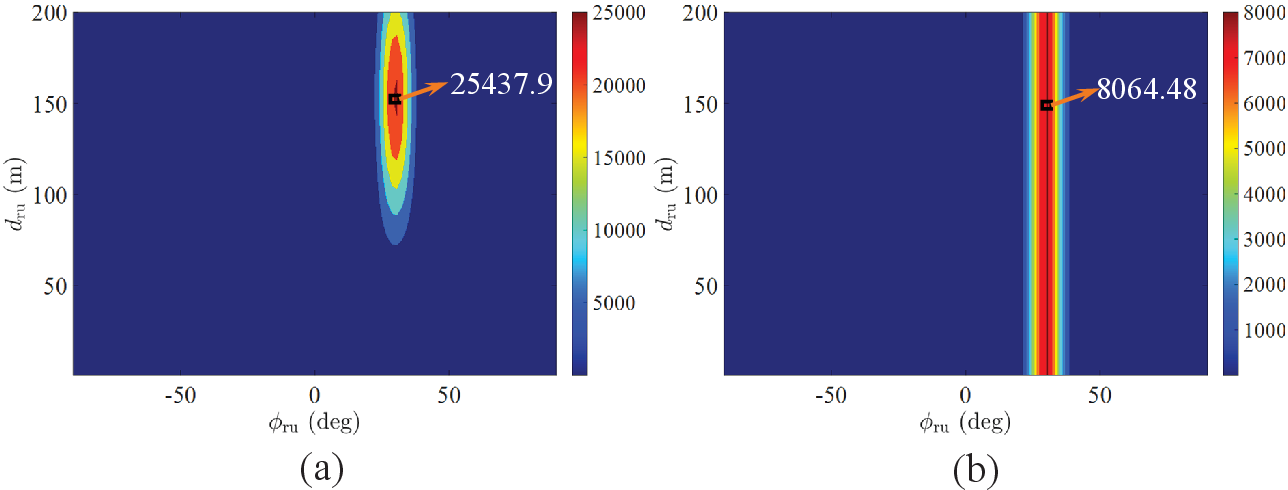}\\
\caption{The generated beam pattern with $S=100$, $P=30$ dBm, target user position ($d_\mathrm{ru}=150$ m, $\theta_\mathrm{ru}=90^{\circ}$ and $\phi_\mathrm{ru}=30^{\circ}$): (a) the beam pattern of the FD-RIS, (b) the beam pattern of the conventional RIS.}\label{fig:pattern}
\end{figure}

Fig. \ref{fig:pattern}(a) and \ref{fig:pattern}(b) display the beam patterns of the FD-RIS and conventional RIS, respectively, considering a target user with position at ($d_\mathrm{ru}=150$ m, $\theta_\mathrm{ru}=90^{\circ}$,  $\phi_\mathrm{ru}=30^{\circ}$). 
Note that the amplitudes in the beam patterns signify the received signal power at various locations, we disregard the path loss effects in order to see the difference between two schemes. Compared with the RIS that can only control the angle dimension as shown in Fig. \ref{fig:pattern}(b), it is easy to observe that the FD-RIS definitely has the capability to control the incident signals from joint distance-angle dimensions and concentrate more signal energy to the target user as shown in Fig. \ref{fig:pattern}(a).
It is worth noting that more than three times of signal energy  can be transmitted to users with the assistance of FD-RIS compared to the result with the assistance of  the RIS  (25327.9 versus 8064.48). The beam pattern results clearly demonstrate that the considerable potential of the FD-RIS.
\begin{figure}[ht]
	\centering
	\begin{minipage}[t]{0.24\textwidth}
		\centering
		\includegraphics[scale=0.23]{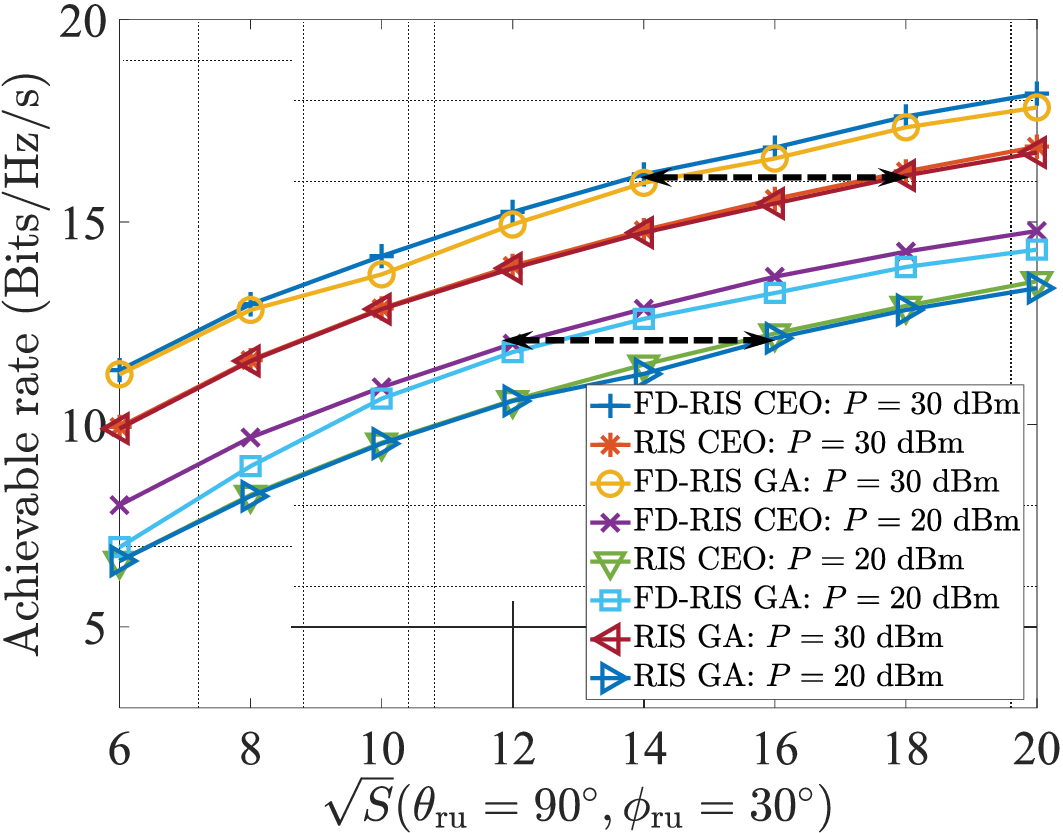}\\
		\caption{The achievable rate versus $S$ with different transmit power ($P$).}\label{fig:S_rate}
	\end{minipage}
	\hfill
\begin{minipage}[t]{0.24\textwidth}
	\centering
	\includegraphics[scale=0.23]{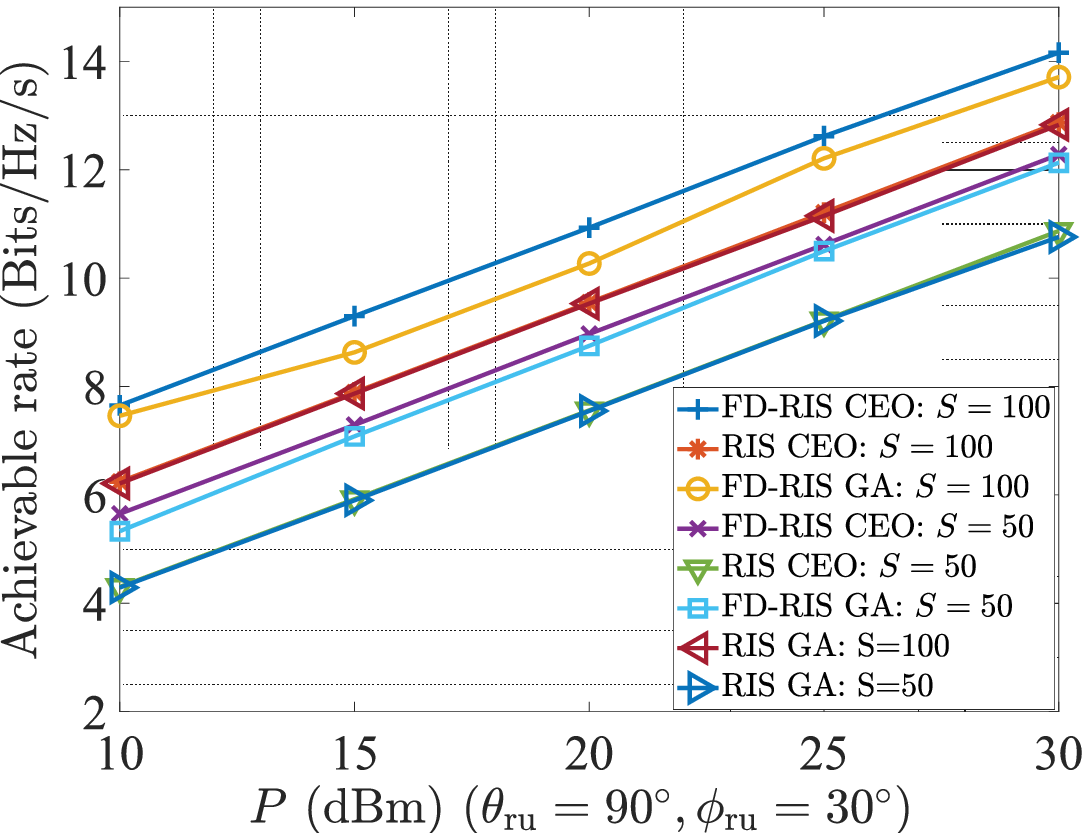}\\
	\caption{The achievable rate versus the transmit power ($P$) with $S$.}\label{fig:P_rate}
\end{minipage}
\end{figure}

The effects of the number of elements $S$ on the achievable rate are investigated considering different the transmitting power and user's locations, as shown in Fig. \ref{fig:S_rate}. Specifically, the attainable rates demonstrate positive correlations with the number of elements deployed on RIS across all scenarios, due to the fact that more elements can provide more degrees of freedom to control beamforming toward the user. Additionally, compared with the GA, which has the probability of finding the global optimal solution, the proposed CEO algorithm demonstrates a slight performance advantage, thereby confirming the effectiveness of the proposed method. \textcolor{blue}{The proposed FD-RIS-assisted scheme exhibits a discernible performance advantage compared with the RIS-aided schemes including RIS CEO scheme and RIS GA scheme (over $1.3$ dB improvement)}. It is worth noting that the FD-RIS can achieve the same performance gain as the RIS even when the number of elements is reduced by more than $35\%$. These underscore the potential of the FD-RIS in augmenting far-field communication performance.
\begin{figure}[ht]
	\centering
	\includegraphics[scale=0.4]{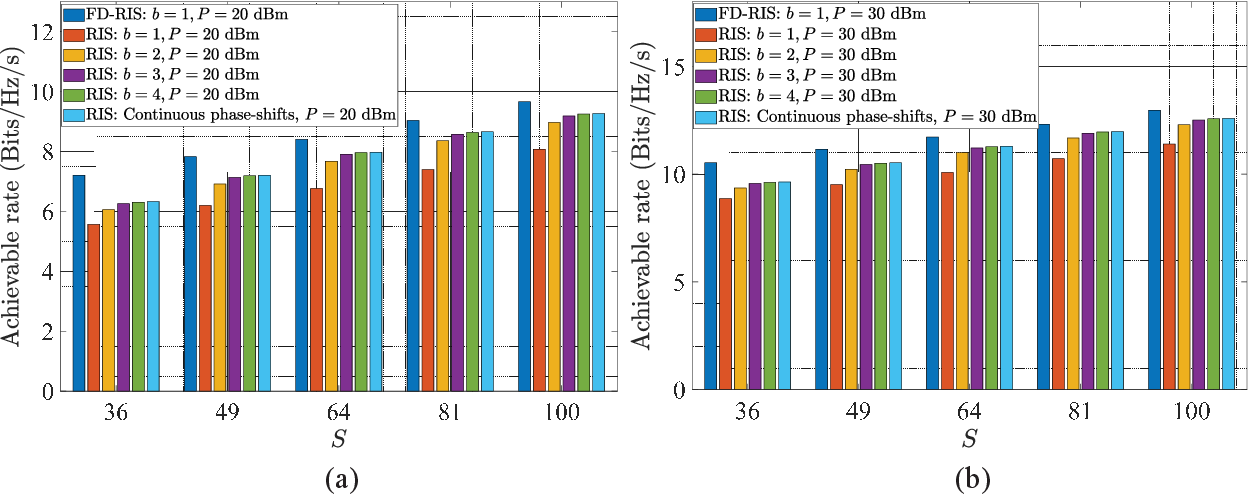}\\
	\caption{The achievable rates of the FD-RIS with $1$-bit and the RIS with different $b$, considering different $S$ and $P$.}\label{fig:FD-RIS_vs_RIS_different_bits}
\end{figure}

In  Fig. \ref{fig:P_rate}, we investigate the varied correlation of the achievable rate with BS's transmit power $P$  considering different $ S$ and target user locations. Specifically, we can find that the attainable rates show an almost linear growth as $P$ rises for all cases, due to the fact that more energy can be used for transmissions. \textcolor{blue}{Furthermore, it is observed that the suggested FD-RIS-aided scheme shows an approximate $1.4$ dB improvement in system capacity when contrasted with the RIS-aided scheme.}

\textcolor{blue}{In Fig. \ref{fig:FD-RIS_vs_RIS_different_bits}, we further compare the achievable rate between the FD-RIS with 1-bit quantization FD-RIS with $b$-bit quantization traditional RIS considering different $S$ and $P$. Specifically, the conventional RIS involves phase-shifts ranging from $1$-bit to $4$-bit as well as continuous phase-shifts. The obtained results indicate that despite employing the lowest precision phase-shift ($1$-bit), the FD-RIS still demonstrates significant performance advantages even when compared to the RIS with continuous phase-shifts. In other words, under equivalent performance constraints, the FD-RIS can satisfy requirements using lower phase-shift accuracy, which greatly facilitates the practical implementation of the FD-RIS.}

\section{Conclusion}\label{sec:S6}
In this paper, we initially design the FD-RIS which can achieve joint distance-angle beamforming  to further enhance the wireless communication performance. Specifically, we first derive the signal processing model of the FD-RIS-supported wireless networks to verify the frequency diverse ability of the FD-RIS. Then, we formulate an optimization problem aimed at maximizing  user's achievable rate by jointly designing the time modulation frequency and reflection coefficients of the FD-RIS. A novel iterative algorithm utilizing the CEO framework is designed to efficiently address the non-convex optimization problem. The simulation results indicate that the joint distance-angle control capability achieved by the FD-RIS has promising potentials to break through the performance limitation of the far-filed communications.

\appendices

\vspace{-1mm}
\ifCLASSOPTIONcaptionsoff
  \newpage
\fi
\bibliographystyle{IEEEtran}
\bibliography{FD-RIS}

\end{document}